\documentstyle[aps,prl,epsf,twocolumn]{revtex}
\begin{document}
\def\bea{\begin{eqnarray}}
\def\eea{\end{eqnarray}}
\def\a{\alpha}
\def\d{\delta}
\def\p{\partial} 
\def\nn{\nonumber}
\def\r{\rho}
\def\xv{\bar{x}}
\def\rv{\bar{r}}
\def\la{\langle}
\def\ra{\rangle}
\def\e{\epsilon}
\def\o{\omega}
\def\n{\eta}
\def\g{\gamma}
\def\th{\hat{t}}
\def\uh{\hat{u}}
\def\break#1{\pagebreak \vspace*{#1}}
\def\f{\frac}
\draft
\title{ Triple minima in Free Energy of Semiflexible Polymers} 
\author{Abhishek Dhar $^1$ and Debasish Chaudhuri $^2$}
\address{  $^1$Raman Research Institute,
Bangalore 560080, India\\ 
$^2$ S.N.Bose National Center for Basic Sciences ,
 Calcutta 700098, India.\\ 
}
%\date{\today}
\maketitle
\widetext
\begin{abstract}
We study the free energy of the
worm-like-chain model, in the constant-extension ensemble, as a function
of the stiffness $\lambda$ for finite chains of length $L$. 
We find that the polymer properties obtained in this ensemble are
{\it qualitatively} different from those obtained using constant-force ensembles. 
In particular we find 
that as we change the stiffness parameter, $t=L/\lambda$, the
polymer makes a transition from the flexible to the rigid phase and there
is an intermediate regime of parameter values where the free energy
has three minima and both phases are stable. This leads to 
interesting features in the force-extension curves.
\end{abstract}

\narrowtext

The simplest model for describing semiflexible polymers without
self-avoidance is the so called Worm-Like-Chain (WLC) model \cite{doi,saito,frey}. In this
model the polymer is modeled as a continuous curve that can be specified by a
$d-$dimensional ($d > 1$) vector $\bar{x}(s)$, $s$ being the distance,
measured along the length of the curve, from one fixed end. The energy of
the WLC model is just the energy due to curvature and is given by 
\bea
\f{H}{k_BT}=\f{\kappa}{2} \int_0^L (\f{\p \uh (s)}{\p s})^2 ds,
\eea
where $\uh (s)=\p{\bar{x}}/\p{s} $ is the tangent vector and satisfies
$\uh^2=1$. The parameter $\kappa$ specifies the stiffness of the
chain and is related to the persistence length $\lambda$ defined
through $\la \uh(s). \uh(s') \ra = e^{-|s-s'|/ \lambda}$. It can
be shown that $\kappa= (d-1) \lambda/2$ .

The thermodynamic properties of such a chain can be obtained from the
free energy which can be either the Helmholtz's $(F)$ free energy or
the Gibb's $(G)$ 
energy. In the former case one considers a polymer whose ends are kept
at a fixed distance $r$ [one end fixed at the origin and the other end
at $\rv =(0,...0,r)$] 
 by an average force $ \la f \ra = \p F(r,L)/\p r$, while in the latter
case one fixes the force and the average extension is given by $\la r \ra= -\p G(f,L)/
\p f$. It can be shown that in the thermodynamic limit $L \to \infty$
the two ensembles are equivalent and related by the usual Legendre 
transform $G=F-fr$.        
For a system with finite $L/\lambda$, the equivalence of the two
ensembles is not guaranteed, especially when fluctuations become
large. We note that real polymers come with a wide range of 
values of the parameter $ t=L/\lambda$ [e.g. $\lambda \approx 0.1 \mu m$ for
DNA while $\lambda \approx 1 \mu m$ for 
Actin and their lengths can be varied] and fluctuations in $r$ (or $f$)
can be very large. Then the choice of the ensemble depends on 
the experimental conditions. Experiments on stretching polymers are usually
performed by fixing one end of the polymer and attaching the other end
to a bead which is then pulled by various means (magnetic, optical,
mechanical, etc.). In such experiments one can either
 fix the force on
the bead and measure the average \break{1.2in} polymer extension, or, one could constrain
the bead's position and look at the average force on the polymer.
In the former case, the Gibb's free energy is relevant while
it is the Helmholtz   in the second case. This point has been carefully
analyzed by Kreuzer and Payne in the context of atomic force microscope
experiments \cite{kreu}. 
Theoretically, the constant-force
ensemble is easier to treat, and infact an exact numerical solution
has been obtained \cite{marko} (though only for $ t >> 1$).  
Data on force-extension experiments on DNA \cite{smith}
have been explained using this ensemble  \cite{marko}.    
The case of constant-extension ensemble turns out to be much harder
and no exact solution is available. The $t \to 
0$ and $t \to \infty $ cases correspond to the solvable limits  of the
hard rod and the Gaussian chain. The small and large $t$ 
cases have been treated analytically  by perturbation theory about
these two limits \cite{dan,gob,nori}. Numerical simulations for
different values of $t$ have been reported by Wilhelm and Frey
\cite{wilh}, who have also obtained series expansions valid in the
small $t$ limit. A mean-field treatment has also recently been
reported \cite{thiru}.

In this letter we probe the nature of the transition from the Gaussian
to the rigid rod with change of stiffness 
as shown by the form of the Helmholtz free energy of the  WLC model
(or equivalently the distribution of end-to-end distance). Extensive
simulations are performed in two and three dimensions using the
equivalence of the WLC 
model to a random walk with one-step memory. We find the surprising
result that, over a range of values of $t$, the free energy has three 
minima. This is verified in a one-dimensional version of the model
which is exactly solvable. 

We first note that the WLC model describes a particle in
$d-$dimensions moving with a constant speed (set to unity) and with a random
acceleration. It is thus described by the propagator
\bea
&& Z(\xv,\uh,L|\xv',\uh',0)
 = \f{ \int^{(\xv,\uh)}_{(\xv',\uh')} {\cal{D}}[\xv(s)] 
e^{-H/k_BT}}{  \int {\cal{D}}[\xv{(s)}]
e^{-H/k_BT} }  
\eea
where in the numerator only paths $\xv(s)$, satisfying
$\xv(0)=\xv',\xv(L)=\xv, \uh(0)=\uh' $ and $ \uh(L)=\uh$ are
considered. It can be shown that the corresponding probability
distribution $W(\xv,\uh,L)$ 
satisfies the following Fokker-Planck equation \cite{dan,gob}:
\bea
\f{\p W}{\p L}+\uh.\nabla_{\xv} W - \f{1}{2 \kappa} \nabla_{\uh}^2 W=0
\label{fkpl}
\eea
where $\nabla_{\uh}^2$ is the diffusion operator on the surface of the
unit sphere in $d-$dimensions. 
The discretized version of this model is the freely
rotating chain model (FRC) of semiflexible polymers \cite{doi}. In the FRC one
considers a polymer with $N$ segments, each of length
$b=L/N$. Successive segments are constrained to be at a fixed angle,
$\theta$, with each other. The WLC model is obtained, in the limit
$\theta , b \to 0$, $ N \to \infty$ keeping $\lambda =2 b/ \theta^2$
and $L=Nb$ finite.

Here we will consider the  situation where the 
ends are kept at a fixed separation $r$ [with $\xv'$ at the origin and
$\xv=\rv=(0,...0,r)$]
but there is no constraint on $\uh$ and 
$\uh'$ and they are taken as uniformly distributed. Thus we will be
interested in the distribution 
$P(r,L)=\la \d (\xv-\rv) \ra= \int d\uh W(\rv,\uh,L) $: this gives
the Helmholtz free energy $F(r,L)=-Log[P(r,L)]$. 
For the spherically symmetric situation we are considering, $P(r,L)$
is simply related to the radial probability distribution $S(r,L)$
through $S(r,L)=Cr^{d-1}P(r,L)$, $C$ being a constant equal to the area of
the $d-$dimensional unit sphere. 
It may be
noted that the WLC Hamiltonian is  equivalent to spin $O(d)$ 
models in one dimension in the limit of the exchange constant $J \to
\infty$ (with $Jb =\kappa$ finite) and 
all results can be translated into spin language.
However, for spin systems, the present free energy is not very relevant since it
corresponds to putting unnatural constraints on the magnetization
vector.

Numerical simulations: The simulations were performed by generating
random configurations of the FRC model and computing the distribution
of end-to-end distances. 
To obtain equivalence with the WLC model the appropriate limits were
taken. We note that 
because these simulations do not require equilibration, they are much
faster than simulations on equivalent spin models and give
better statistics. The number of configurations generated was 
around $10^8$ for  chains of size $N=10^3$. We verified that
increasing $N$ did not change the data significantly. As a
check on our numerics we evaluated $\la r^2 \ra $ and $\la r^4
\ra$. Using Eq.~(\ref{fkpl}) and following \cite{herm} we can compute these (in all dimensions):
\bea
&&\la r^2 \ra=\f{4 \kappa L}{d-1}-\f{8 \kappa^2(1-e^{-\f{(d-1)L}{2
\kappa}})}{(d-1)^2}   \nn \\ 
&& \la r^4 \ra =\f{64 \kappa^4(d-1)}{d^3 (d+1)^2}e^{-\f{dL}{\kappa}}-\f{128 \kappa^4
(d+5)^2}{(d-1)^4 (d+1)^2}e^{-\f{(d-1)L}{2\kappa}} +\nn \\ 
&& \f{64 \kappa^3 L
(d^2-8 d+7)}{(d-1)^4(d+1)} e^{-\f{(d-1)L}{2 \kappa}}+\f{64 \kappa^4 (d^3+23 d^2-7d+1)}{(d-1)^4 d^3} \nn \\ 
&& -\f{64 \kappa^3 L
(d^3+5 d^2-7d+1)}{(d-1)^4 d^2}+\f{16 \kappa^2 L^2 (d^3-3d+2)}{d(d-1)^4}.
\eea
Infact it is straightforward to compute all even moments, though it
becomes increasingly tedious to get the higher moments. 
Our numerics agrees with the exact results to around
$0.1 \% $ for $\la r^2 \ra$ and $0.5\%$ for $\la r^4 \ra$. 
\vbox{
\vspace{0.5cm}
\epsfxsize=7.0cm
\epsfysize=6.0cm
\epsffile{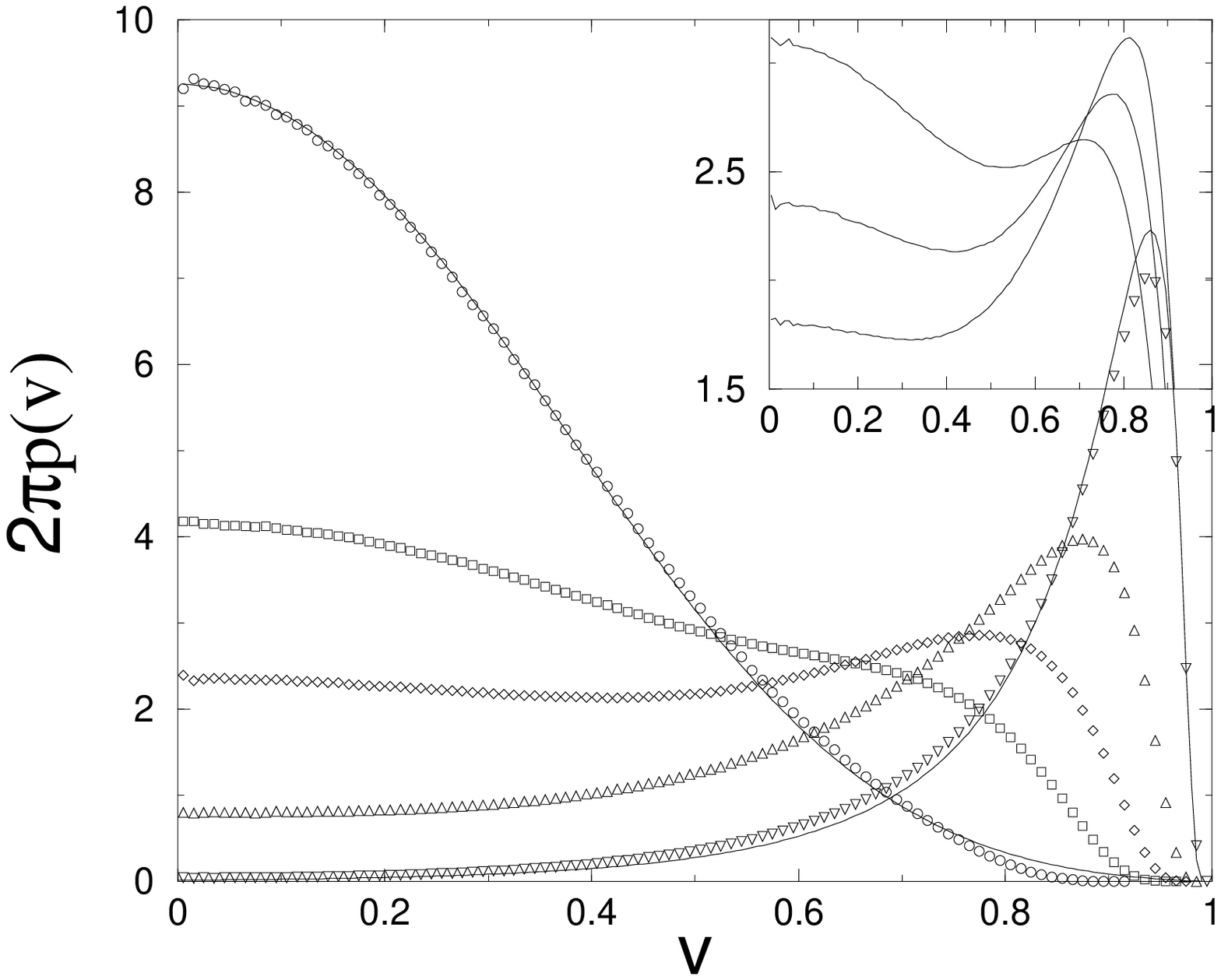}
\begin{figure}
\caption{ Monte-Carlo data for $p(v,t)$ for the $2-$dimensional WLC for
values of $t=10 (\circ),5,3.33,2$ and $1 (\nabla)$. The inset is a
blowup of curves in the transition region ($t=4,3.33,2.86$) and clearly shows the
presence of the two maxima. Note that because of $\pm v$ symmetry, we
have plotted data for positive $v$ values only. For the fits at large
and small $t$ see text.     
\label{wlc2d}
}
\end{figure}}
\vbox{
%\vspace{0.5cm}
\epsfxsize=7.0cm
\epsfysize=6.0cm
\epsffile{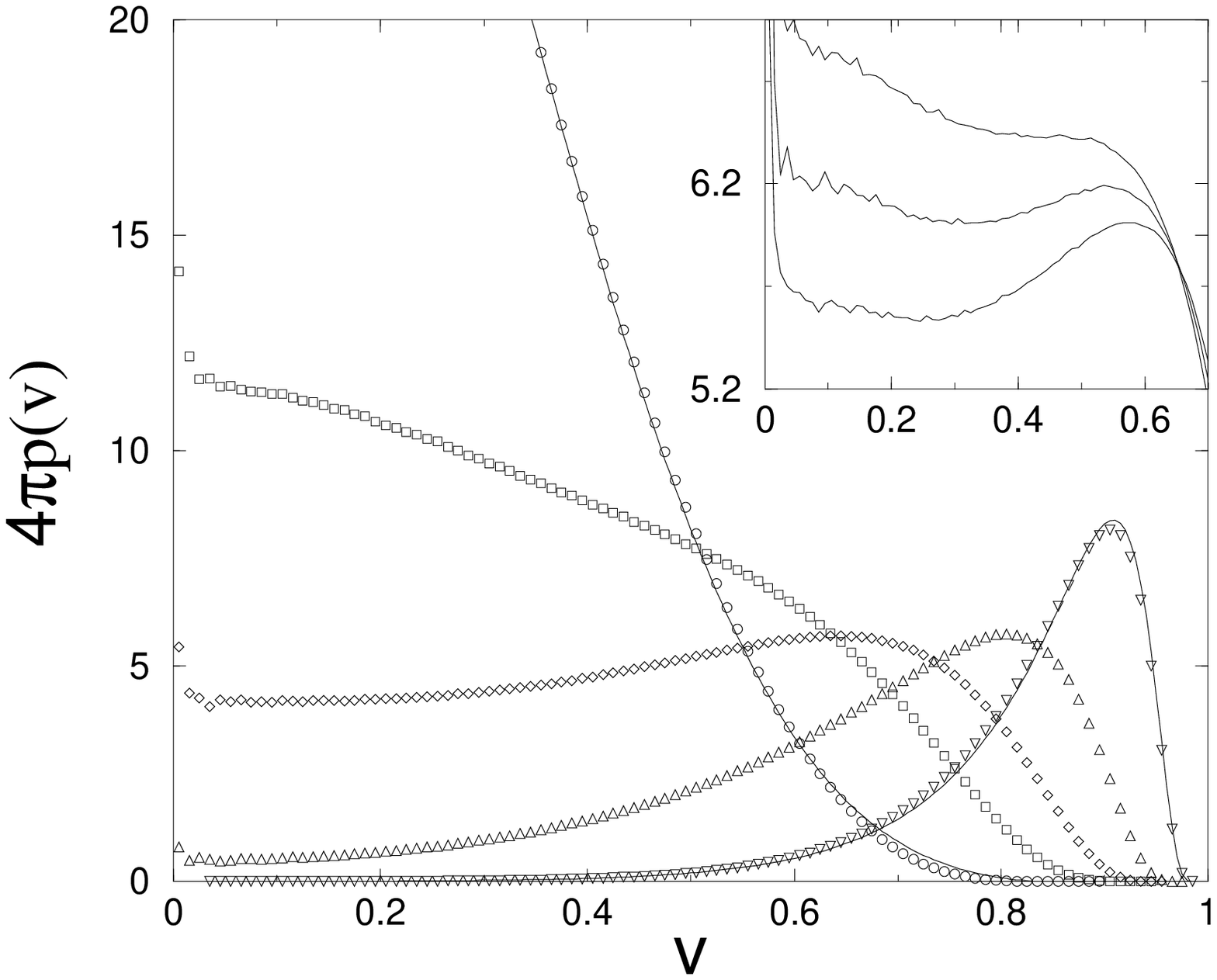}
\begin{figure}
\caption{ Monte-Carlo data for $p(v,t)$ for the $3-$dimensional WLC for
values of $t=10 (\circ),5,3.33,2$ and $1 (\nabla)$. The inset is a
blowup of curves in the transition region ($t=4,3.85,3.7$) and shows the
presence of the two maxima. 
\label{wlc3d}
}
\end{figure}}

The function $P$ has the scaling form $P(r,L)=\f{1}{L^d} p(r/L,L/\lambda )$ and we will
focus on determining the function $p(v,t)$ \cite{ftnt}. 
In Fig.~(\ref{wlc2d}) and Fig.~(\ref{wlc3d}), we show the results of
our simulations in two and three dimensions. 
At large values of $t$ there is a
single maximum at $v=r/L=0$ corresponding to a Gaussian distribution while at
small $t$, the maximum is close to the fully extended value of
$v=\pm 1$. The transition is {\it first-order-like}: as we decrease $t$,
at some critical value, $p$ develops two additional
maxima at non-zero values of $v$. Further decreasing $t$ weakens the
maximum at $v=0$ until it finally disappears and there are
just two maxima which correspond to the rigid chain.  

For the limiting cases of small and large values of $t$ there are
analytic results for the distribution function and as can be seen in
Fig.~(\ref{wlc2d},\ref{wlc3d}) our data agrees with them.
 For large $t$ we find that Daniels approximation \cite{dan},
which is a perturbation about the Gaussian, fits the data quite
well. In the other limit of small $t$ the series solutions provided in
\cite{wilh} fits our data. 
For intermediate values of $t$ neither of the two forms are able to
capture, even qualitatively, the features of the free energy. 
Specifically, we note that all the analytic theories(perturbative, series
expansions and mean-field)  
predict a second-order-like
transition and do not give triple minima of the free energy for any
parameter value.

It is instructive to study a one-dimensional version of the WLC
which shows the same qualitative features (the equivalent spin problem
is the Ising model). Consider a $N$ step random walk,
with step-size $b$ which, with probability 
$\epsilon$, reverses its direction of motion and with $1-\epsilon$,
continues to move in the same direction. 
The appropriate scaling limit is: $b \to 0$, $\epsilon
\to 0$, $N \to \infty$ keeping $L=Nb$,  $t=L/\lambda =2N\epsilon $ finite.
Defining $Z_{\pm}(x,L)$ as the probability of the walker to be at $x$
with either positive or negative velocity, we have the following
Fokker-Planck equation:
\bea
\f{\p Z_{\pm}}{\p L}= \mp \f{\p Z_{\pm}}{\p x} \mp \f{1}{2 \lambda}
(Z_+-Z_-)
\eea    
This can be solved for $P(x,L)=Z_++Z_-=\f{1}{L} p(x/L,L/\lambda)$. We get
\bea
&& p(v,t)=  \f{te^{-t/2}}{4}
[\f{I_1(\f{t}{2}\sqrt{1-v^2})}{\sqrt{1-v^2}}
+I_0(\f{t}{2}\sqrt{1-v^2})] \nn \\
&&~~~~~~~~~  + \f{e^{-t/2}}{2} [\d (v-1) + \d (v+1) ],
\label{p1d}
\eea
where $I_0$ and $I_1$ are modified Bessel functions. 
In Fig.~\ref{wlc1d}
we  have plotted $p(v,t)$ for different 
values of stiffness. We find that it always has three
peaks. Unlike in $2$ and $3$ dimensions, the
$\delta-$function peaks at $v=\pm1$ (which corresponds to fully
extended chains) persist at all values of stiffness though their
weight decays exponentially. Similarly the peak at $v=0$ is always present.

Discussion: The most interesting result of this paper is the triple
minima in the Helmholtz free energy of the WLC. 
Physically, this results from the competing effects of
entropy, which tries to pull in the polymer and 
the bending energy,
which tries to extend it. 
This form of the free energy 
leads to a highly {\it counterintuitive force-extension curve},  
very different from what one obtains from the
constant force ensemble or from approximate theories. In
Fig.~(\ref{fext}) we show the 
force-extension curve for a two dimensional chain with $t=3.33$. We
see that there are two stable positions for which the force is zero.   
In the constant-force ensemble, it is easy
to show that $\p \la r \ra /\p f=\la r^2 \ra-\la r \ra^2$ and so the
force-extension is always monotonic. In the
constant-extension ensemble, there 
is no analogous result ({\it{for finite systems}}), and monotonicity is not
guaranteed.  
\vbox{
%\vspace{0.5cm}
\epsfxsize=7.0cm
\epsfysize=6.0cm
\epsffile{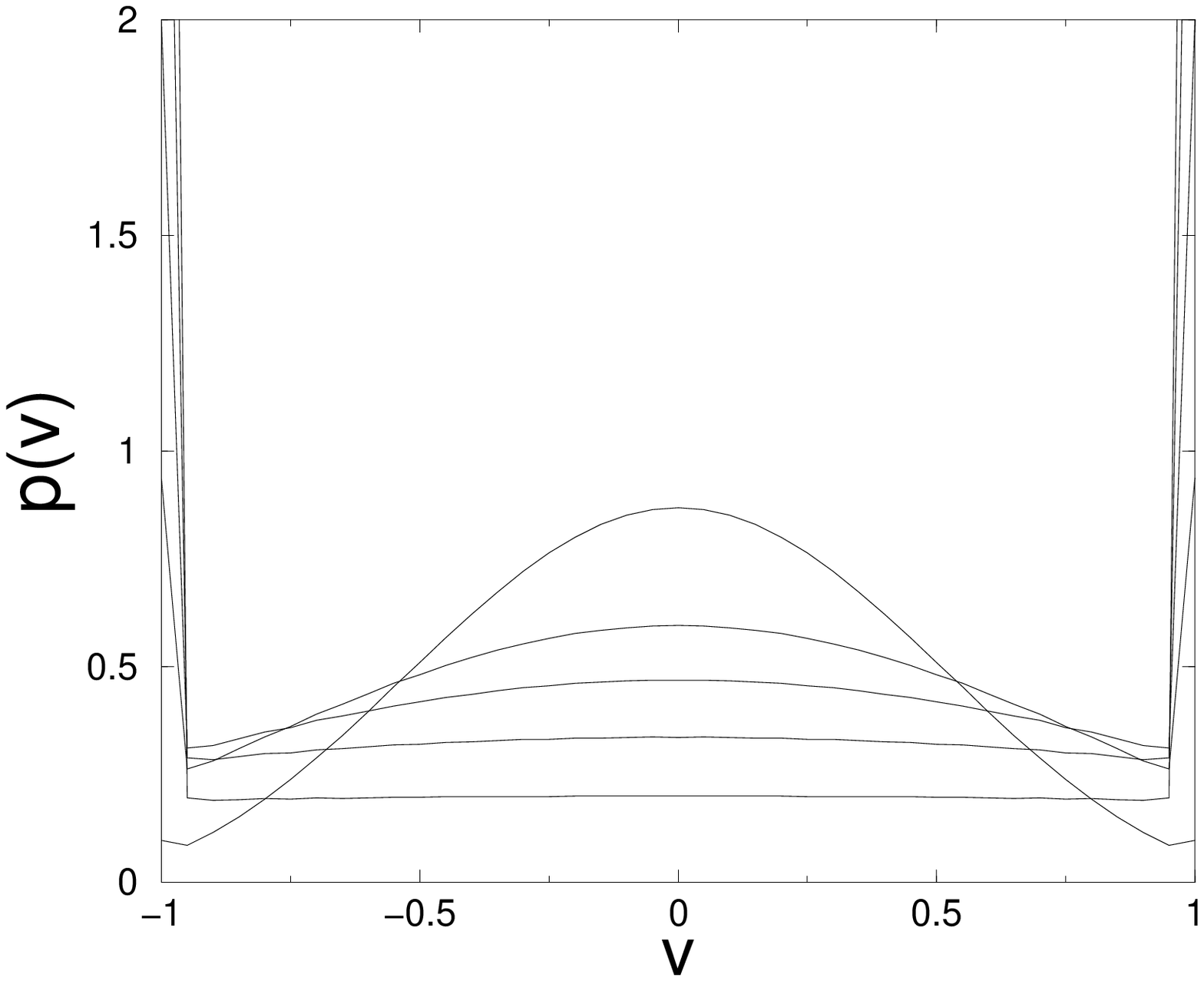}
\begin{figure}
\caption{ The exact distribution $p(v,t)$ of the $1-$dimensional WLC
[ Eq.~(\ref{p1d})] for different
values of $t$ ($10,5,3.33,2,1$). Even for the most stiff chain
considered here ($t=1$), the distribution has a peak at the centre (in
addition to the $\delta-$function peaks at ends) though it looks flat.
\label{wlc1d}
}
\end{figure}}

Most of the recent experiments on stretching DNA have  $t
\gtrsim 100 $. The distribution is 
then sharply peaked at
zero and one expects the equivalence of different ensembles.    
Experimentally the value of $t$ can be tuned by
various means, for example, 
by changing the length of the polymer or
the temperature. 
Polymer-stretching experiments can thus be performed
for intermediate  $t$ values.
Since we consider the tangent vectors at the polymer-ends to
be unconstrained an accurate experimental realization of our set-up
would be one in which both ends are attached to beads [see
Fig.~(\ref{schm})]. The beads are
put in optical traps and so are free to rotate (this setup is
identical to the one used in refn.~\cite{mein}). Making the traps stiff
corresponds to working in the constant-extension ensemble \cite{kreu} and one can
measure the average force.  Our predictions can then be experimentally
verified. We make some estimates on the experimental
requirements (for a $3$-d polymer with stiffness $t=3.85$). 
Assume that at one end, the origin, the trap is so stiff that 
the bead can only rotate. We make measurements at the other end. The
trap-center is placed at $\rv_0=(0,0,z_0)$ and the mean bead displacement
$\Delta z= \la (z-z_0) \ra $ gives the mean force $\la f \ra$ on the polymer. 
 We then consider the problem of the polymer in the presence of a  trap
potential $V=[k_t(x^2+y^2)+k (z-z_0)^2]/2$. Assume $k_t >> k$ so we
can neglect fluctuations in the transverse directions. The 
distribution of the bead's position in the presence of 
the potential is given by 
\vbox{
%\vspace{0.25cm}
\epsfxsize=7.0cm
\epsfysize=6.0cm
\epsffile{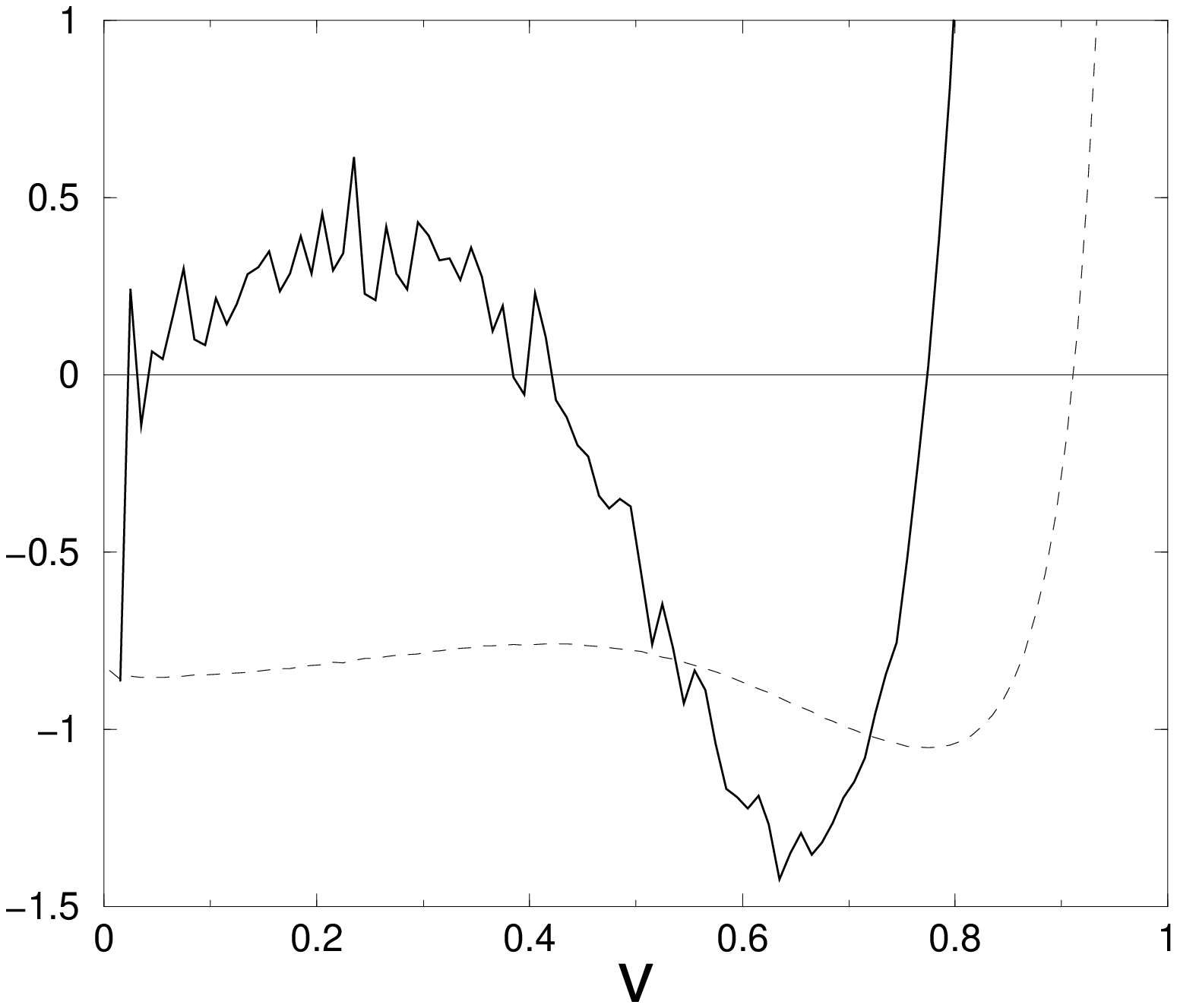}
\begin{figure}
\caption{ The free energy (dotted line) and the corresponding
force-extension curve (solid line) for a $2-$dimensional chain with $t=3.33$.
\label{fext}
}
\end{figure}}
\vbox{
%\vspace{0.5cm}
\epsfxsize=7.0cm
\epsfysize=5.0cm
\epsffile{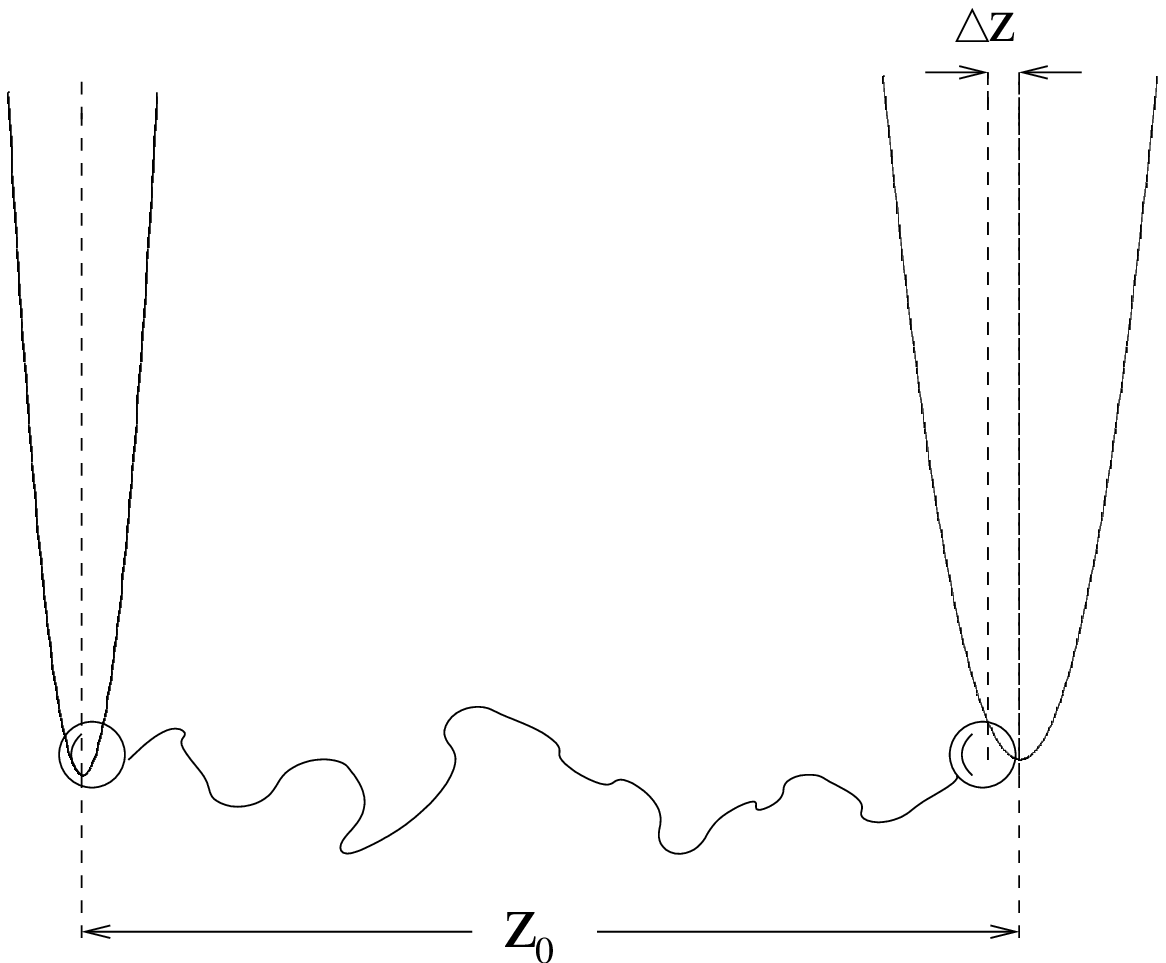}
\begin{figure}
\caption{ 
A schematic of the experimental set-up required to realize the
constant-extension ensemble discussed in the paper (see
refn. [14]). For a stiff trap the average 
displacement of the bead $\la \Delta z \ra $ from the trap center is
small and the average force on the polymer is: $\la f \ra =-k \la \Delta z
\ra $. 
\label{schm}}
\end{figure}}
$Q(\rv)=e^{-\beta [F(\rv)+V(\rv)]}/\int d^3
\rv e^{-\beta [F(\rv)+V(\rv)]}$. For a stiff trap, we can expand $F$
about $\rv=\rv_0$ and find that the average displacement
of the bead is given by: $\Delta z= \int d^3 \rv
(z-z_0) Q(\rv)= -\la f \ra/k'$, where $k'=k+F''(z_0) \approx k$ (valid except when
$z_0 \approx L$) and $\la f \ra=F'(z_0)$. The rms fluctuation of the
bead about the trap center is given by 
$z^2_{rms}=k_B T /k$. Hence we get $  \Delta z = -\la f\ra z_{rms}^2/(k_B T)=
-(\la f\ra L/k_B T) (z_{rms}^2/L)$. The scaled force $\la f \ra L/(k_B T)$ is of order
$0.1$. The different minima are separated by  distances $\approx 0.2 L$, hence to
see the effect we need to have $z_{rms}/L \lesssim 0.1$. Thus finally
we find that the typical displacement of the bead $\Delta z$ is about  $0.01
z_{rms}$. This is quite small and means that 
it is necessary to collect data on the bead position over long periods of
time.  

As suggested in \cite{wilh}, a more direct way of measuring the
Helmholtz free energy would be to attach marker molecules at the ends of the
polymer and determine the distribution of end-to-end
distances. Fluorescence microscopy as in \cite{ott} could be another
possible method. 
It is to be remembered of course that real polymers are well-modeled
by the WLC model provided we can neglect monomer-monomer interactions
(steric, electrolytic etc.).
Thus the experiments would really test the relevance of the WLC model in
describing real semiflexible polymers in different stiffness regimes.

In conclusion we have presented some new and interesting properties of
the WLC model and have pointed out that polymer
properties are ensemble-dependent. In this paper we have given one
example of qualitative differences in force-extension measurements in
different ensembles. Other quantitative differences will occur even in
more flexible chains and should be easier to observe
experimentally. We hope this work will motivate further 
experimental and theoretical work on this simplest model for
semiflexible polymers.

We thank O. Narayan and J. Samuel for discussions. 
One of us (D.C.) thanks CSIR, India for support.

Note: After submission of this paper, an exact numerical solution of the WLC
model has been obtained, and has reproduced our results \cite{sam}.


\begin{references}
\bibitem{ftnt} Wilhelm and Frey \cite{wilh} have looked at the radial distribution
$S(r,L)$ which however misses the interesting details of the
transition. We note that the relevant distribution here is indeed
$p(v,t)$ since this gives the Helmholtz free energy.
\end{references}
\end{document}